\documentclass{article}
\usepackage{authblk}

\usepackage[utf8]{inputenc}
\usepackage{siunitx}
\usepackage{graphicx}
\usepackage{amsmath,amsfonts,amssymb}

\newenvironment{keyword}
{\noindent\textbf{Keywords:} }
{\par}

\usepackage[backend=biber,sorting=none]{biblatex}
\usepackage{hyperref}
	
\begin{document}
	
\title{MPC for nonlinear systems: a comparative review of discretization methods}

\author[1]{Guido Sanchez}
\author[1]{Marina Murillo}
\author[1]{Lucas Genzelis}
\author[1]{Nestor Deniz}
\author[1]{Leonardo Giovanini}

\affil[1]{Research Institute for Signals, Systems and Computational Intelligence, sinc(i), FICH-UNL/CONICET, Ciudad Universitaria UNL, $4^\circ$ piso FICH, (S3000) Santa Fe, Argentina.}

\date{September 2017}

		\maketitle
		
		\begin{abstract}
			This work provides a comparative review of three different numerical methods generally used to discretize continuous-time non-linear equations appearing in model predictive control problems: direct multiple shooting, direct collocation and successive linearizations. An overview of the characteristics of each method is given and the performance of each method is evaluated through the simulation of two test cases.
		\end{abstract}


		\section{Introduction}
		
		In this work we review some of the most popular methods used to discretize non-linear continuous-time equations that describe the system dynamics in model predictive control (MPC) algorithms.
		
		MPC is a control technique that uses models of the plant in order to predict the future behaviour of the system over a prediction horizon. Usually, these models are expressed mathematically as ordinary differential equations (ODE). As MPC is, in general, a discrete-time control technique, these ODEs need to be transformed into discrete-time difference equations using a proper discretization method. 
		
		The work by Rao \cite{Rao2009} describes the most important methods that have been developed over the years for solving general optimal control problems, in particular, the two broad classes of indirect and direct methods are discussed. Diehl et al. \cite{diehl2006fast} review numerical approaches to solve non-linear optimal control problems, and discuss three direct optimal control approaches in detail: (i) single shooting, (ii) collocation, and (iii) multiple shooting. 
		Probably, the most commonly used method to discretize continuous-time non-linear equations is the direct single-shooting method due to its simplicity and because it works well for simple problems. In this method the controls are discretized and then a numerical routine is used to sequentially obtain the states as an initial value problem for the complete horizon. Single shooting is very simple; however, when the systems are highly non-linear or unstable it is known to cause problems \cite{kirches2011fast}.
		
		In order to improve direct single shooting, multiple shooting adds more degrees of freedom and partitions the time interval into additional grid points, leading to the generation of many initial value problems, which improves the convergence and prevents the growth of the error introduced by poor initial data \cite{bock1984multiple,morrison1962multiple}.
		
		Another method that works well is direct collocation, which add a series of \textit{collocation points} and approximates the model equations using a polynomial basis. The method ensures that the model equation is satisfied on this intermediate points \cite{ahlberg1975collocation,topputo2014survey}. The addition of these variables adds even more degrees of freedom than multiple shooting, but we are rewarded with less nonlinearity.
		
		An alternative discretization method is successive linearization, which was presented by the authors in \cite{Murillo2015,Murillo2016120}. With this discretization method the continuous-time system equation can be transformed into a set of discretized linear time-varying (LTV) equations. 
		
		In this work we mainly revisit the last three methods: i) direct multiple shooting, ii) direct collocation, and  iii) successive linearization. This work is organized as follows: in Section \ref{prob_form} the problem formulation is presented. In Section \ref{meth_over} an overview of the three methods is given. In order to compare the discretization methods, two test simulation examples are given in Section \ref{examp}. Finally, in Section \ref{concl} the conclusion of this work is stated.

		\section{Problem formulation}
		\label{prob_form}
		In the following we consider continuous-time dynamical systems described by the following equation
		\begin{equation}
			\dot{x}(t) = f(x(t),u(t)),
			\label{dynamical_system}
		\end{equation}
		where $x(t) \in \Re^{n_x}$ is the differential state and $u(t) \in \Re^{n_u}$ is the control. Function $f$ is assumed to be twice differentiable and maps into $\Re^{n_x}$. Given the dynamical system from Eq. (\ref{dynamical_system}), the MPC problem is defined as follows: determine the state $x_k \in \Re^{n_x}$ and the control $u_k \in \Re^{n_u}$ that solves the following optimization problem
		\begin{equation}
			\begin{array}{rrclcl}
				\displaystyle \min_{(x,u)} & \multicolumn{3}{l}{\displaystyle \sum_{k=i}^{i+N-1} \ell_k(x_{k}, u_{k}) + \ell_N(x_{k+N})} \\
				\textrm{s.t.} & x_i - \bar{x}_i &=& 0, \\
				& {x}_{k+1} - f_k(x_k,u_k) &=& 0, \\
				& h_k(x_k,u_k) &\leq& 0, \\
				& \forall k &\in& [i,\ldots,i+N-1], \\
			\end{array}
			\label{nmpc_nlp}
		\end{equation}
		where $N$ is the control horizon, $\ell_k$ is the stage cost and $\ell_N$ is the terminal cost (a popular choice is the use of quadratic costs \cite{maciejowski2002predictive}). The parameter $\bar{x}_i$ is the fixed initial value, the function $f_k$ is the discrete version of Eq. (\ref{dynamical_system}), and the function $h_k$ is the set of state and control constraints.
		In order to solve Problem (\ref{nmpc_nlp}) we need to find a method to discretize Eq. (\ref{dynamical_system}), thus allowing to solve a non-linear program (NLP) on the free variables $(x,u)$. 
		
		\section{Discretization methods overview}
		\label{meth_over}
		In this work we solve the MPC problem using numerical optimization methods, therefore the continuous-time dynamical system equations need to be discretized. During the analysis of the different methods, we will assume a piecewise constant control discretization
		\begin{equation}
			u(t) = u_k, \; \mathrm{for}\; t \in [t_k,t_{k+1}],\; k=i,\ldots,i+N-1,
			\label{control_discretization}
		\end{equation}
		so we can focus on how the state trajectory discretization is handled.

		\subsection{Multiple shooting}
		
		The direct multiple shooting method is due to Bock and Plitt \cite{bock1984multiple} and proceeds as follows: first, the controls are discretized in a piecewise manner on a coarse grid as specified in Eq. (\ref{control_discretization}). Then, each state $x_{k}$ is used inside the grid given by the discrete-time control horizon as a \textit{shooting node} to solve the ordinary differential equation (ODE) given by Eq. (\ref{dynamical_system}) on each interval $[t_k, t_{k+1}]$, starting with the initial state $\bar{x}_i$. From the numerical solution of these initial value problems, we obtain the state trajectory
		\begin{equation}
			{x}_{k+1} = \Phi_k(x_k,u_k),
		\end{equation}
		where $\Phi_k$ is a suitable integrator. 
		Typically, $\Phi_k$ is a Runge Kutta 4 (RK4) approximation which is employed to solve each of the interval's ODE, where the user picks a step size $h>0$ --usually equal to the sampling time-- and then:
		\begin{equation}
			\begin{aligned}
				k_{1} &= f(x_{k}, u_k), \\
				k_{2} &= f(x_{k} + \frac{h}{2} k_1, u_k), \\
				k_{3} &= f(x_{k} + \frac{h}{2} k_2, u_k), \\
				k_{4} &= f(x_{k} + h k_3, u_k), \\
				x_{k+1} &= x_k + \frac{h}{6}(k_{1}+2k_{2}+2k_{3}+k_{4}). \\
			\end{aligned}
		\end{equation}
		The RK4 method is reasonably simple and robust and is a good general candidate for numerical solution of differential equations.
		
		\subsection{Direct Collocation}
		
		While multiple shooting only includes the state at the beginning of each control interval as a degree of freedom in the NLP (in addition to the discretized control and the parameters), in direct collocation the state at a set of \textit{collocation points} (in addition to the beginning of the interval) are variables in the NLP. An example of these time points are the Legendre points of order $d=3$:
		\begin{equation}
			\tau = [0,0.112702,0.500000,0.887298].
		\end{equation}
		Keeping the same control discretization scheme, the complete list of time points inside the control horizon $N$, with $h_k := t_{k+1}-t_k$, is:
		\begin{equation}
			\begin{aligned}
				t_{k,j} &:= t_k + h_k   \tau_j, \\
				k &=i,\ldots,i+N-1, \\
				j&=0,\ldots,d,
			\end{aligned}
		\end{equation}
		as well as the final time $ t_{N,0}$. Also let $ x_{k,j}$ denote the states at these time points.
		On each control interval, we shall define a Lagrangian polynomial basis:
		\begin{equation}
			L_j(\tau) = \prod_{r=0,   r \ne j}^{d} \frac{\tau - \tau_{r}}{\tau_j - \tau_r}.
		\end{equation}
		
		Since the Lagrangian basis satisfies:
		\begin{equation}
			L_j(\tau_r) = \left\{ \begin{array}{l} 1, \qquad \text{if $j=r$} \\ 0, \qquad \text{otherwise} \end{array} \right.,
		\end{equation}
		we can approximate the state trajectory approximation as a linear combination of these basis functions:
		\begin{equation}
			\tilde{x}_k(t) = \sum_{r=0}^{d}{L_r\left(\frac{t-t_k}{h_k}\right)   x_{k,r}}.
		\end{equation}
		In particular, we get approximations of the time derivative of the state at each collocation point (not including $ \tau_0$):
		\begin{equation}
			\displaystyle \tilde{\dot{x}}_k(t_{k,j}) = \frac{1}{h_k}   \sum_{r=0}^{d}{\dot{L}_r(\tau_j)   x_{k,r}} := \frac{1}{h_k}   \sum_{r=0}^{d}{C_{r,j}   x_{k,r}}
			\label{colloc_state_derivative}
		\end{equation}
		as well as an approximation of the state at the end of the control interval:
		\begin{equation}
			\tilde{x}_{k+1,0} = \sum_{r=0}^{d}{L_r(1)   x_{k,r}} := \sum_{r=0}^{d}{D_r   x_{k,r}}.
			\label{colloc_end_state}
		\end{equation}
		
		Plugging the approximation of the state derivative given by Eq. (\ref{colloc_state_derivative}) into the ODE gives us a set of collocation equations that needs to be satisfied for every state at every collocation point:
		\begin{equation}
			\begin{aligned}
				h_k   f(t_{k,j},x_{k,j},u_k) - \sum_{r=0}^{d}{C_{r,j}   x_{k,r}} &= 0, \\
				k&=i,\ldots,i+N-1, \\
				j&=1,\ldots,d. \\
			\end{aligned}
			\label{colloc_ode}
		\end{equation}
		Then, the approximation of the end state given by Eq. (\ref{colloc_end_state}) gives us a set of continuity equations that must be satisfied for every control interval:
		\begin{equation}
			\begin{aligned}
				x_{k+1,0} - \sum_{r=0}^{d}{D_r   x_{k,r}} &= 0, \\
				k &=i,\ldots,i+N-1.
			\end{aligned}
			\label{colloc_continuity}
		\end{equation}
		The set of Eq. (\ref{colloc_ode}) and (\ref{colloc_continuity}) take the place of the integrator call that was made in direct multiple shooting. In direct collocation, simulation and optimization proceed simultaneously. The advantages of collocation methods are (i) a very sparse NLP is obtained (ii) we can use knowledge of the state trajectory in the initialization (iii) it shows fast local convergence (iv) it can treat unstable systems well, and (v) it can easily cope with state and terminal constraints. Its major disadvantage is that adaptive discretization error control needs regridding and thus changes the NLP dimensions \cite{Diehl2009}.
		
		\subsection{Successive linearization}
		
		The method of successive linearizations is based on the work of Murillo et al. \cite{Murillo2015,Murillo2016120}. The main idea is that the dynamic behavior of the non-linear system of Eq. (\ref{dynamical_system}) can be approximated as a linear time-varying (LTV) model
		\begin{equation}
			{x}_{k+1|k} = \hat{A}_{k|k} {x}_{k|k} + \hat{B}_{u_{k|k}} {u}_{k|k},
			\label{ltv1}
		\end{equation}
		where the matrices $\hat{A}_{k|k}$ and $\hat{B}_{u_{k|k}}$ are discretized versions of the Jacobian matrices ${A}_{k|k}$ and ${B}_{u_{k|k}}$ of the continuous time non-linear system given by Eq. (\ref{dynamical_system}), and they are defined as follows
		\begin{equation}
			\begin{array}{rl}
				A_{k|k} &= \frac{\partial f(x_k,u_k,d_k)}{\partial x_k}\biggr\rvert_{(\ast)}, \\
				B_{u_{k|k}} &= \frac{\partial f(x_k,u_k,d_k)}{\partial u(k)}\biggr\rvert_{(\ast)},
			\end{array}
			\label{ss_cont_mat}
		\end{equation}
		where $(\ast)$ stands for the reference trajectory given to the states and controls. The matrices obtained from Eq. (\ref{ss_cont_mat}), must then be discretized using a proper method \cite{decarlo1989linear}.
		As a result, the optimization Problem (\ref{nmpc_nlp}) can be rewritten as follows:
		\begin{equation}
			\begin{array}{rrclcl}
				\displaystyle \min_{(x,u)} & \multicolumn{3}{l}{\displaystyle \sum_{k=i}^{i+N-1} \ell_k(x_{k}, u_{k}) + \ell_N(x_{k+N})} \\
				\textrm{s.t.} & x_i - \bar{x}_i &=& 0, \\
				& {x}_{k+1|k} &=& \hat{A}_{k|k} {x}_{k|k} + \hat{B}_u{_{k|k}} {u}_{k|k}, \\
				& h_k(x_k,u_k) &\leq& 0, \\
				& \forall k &\in& [i,\ldots,i+N-1], \\
			\end{array}
			\label{nmpc_ltv}
		\end{equation}
		and thus can be solved using quadratic program (QP) solvers instead of NLP solvers, given that the set of constraints $h_k$ are linear.
		
		\section{Examples}
		In the following examples we will assume that no perturbations are present and the model used by the controller matches the plant. We will minimize a quadratic cost, so the terms $\ell_k$ and $\ell_N$ will be defined as
		\begin{equation}
			\begin{array}{rcl}
				\ell_k &=& (x_k-x_k^s)^T Q (x_k-x_k^s) + (u_k-u_k^s)^T R (u_k-u_k^s)  \\
				\ell_N &=& x_{k+N}^T Q_N x_{k+N}  \\
			\end{array}
		\end{equation}
		where the matrices $Q \in \Re^{n_x \times n_x}$, $R \in \Re^{n_u \times n_u}$ and $Q_N \in \Re^{n_x \times n_x}$ are symmetric definite positive and $x_k^s$ and $u_k^s$ are the desired state and control setpoints.
		
		\label{examp}
		\subsection{First example: Van der Pol oscillator}
\begin{figure}
			\centering
			\includegraphics[width=1.0\linewidth]{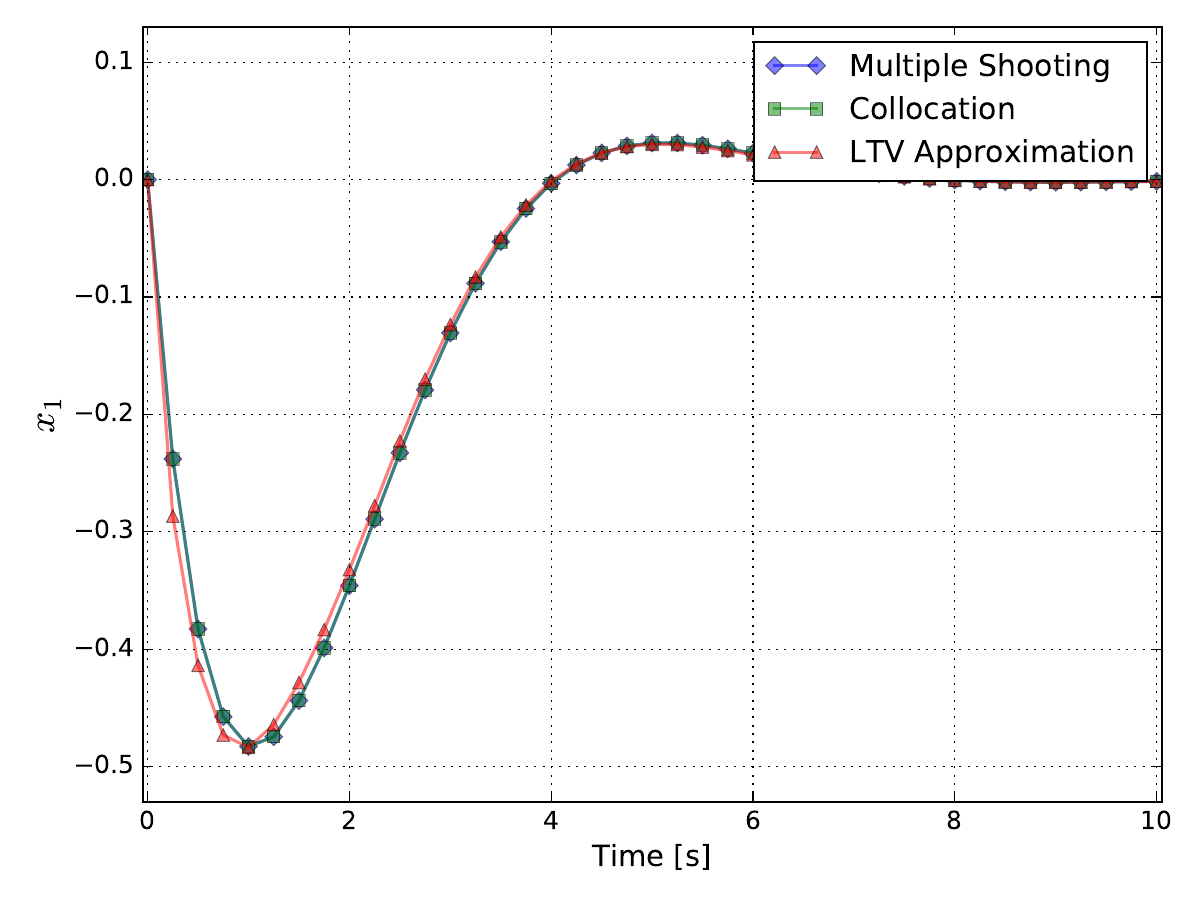}
			\caption{State trajectory $x_1$ for the Van der Pol oscillator.}
			\label{vdp_x1}
		\end{figure}

\begin{figure}
			\centering
			\includegraphics[width=1.0\linewidth]{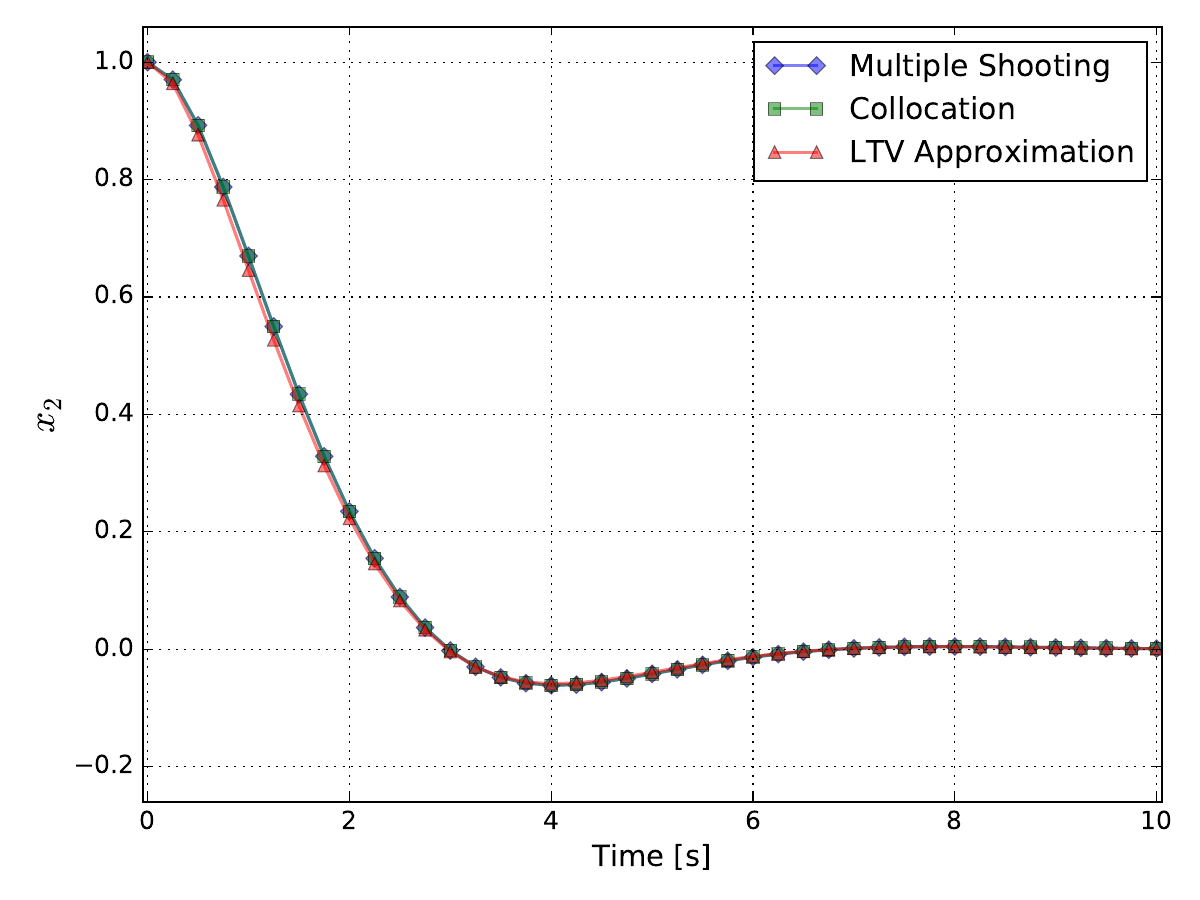}
			\caption{State trajectory $x_2$ for the Van der Pol oscillator.}
			\label{vdp_x2}
		\end{figure}

		First, we will be driving a Van der Pol oscillator to the origin. The system equation is given by
		\begin{equation}
			\begin{array}{rl}
				\dot{x}_1 &= (1-x_2^2)x_1 - x_2 + u, \\
				\dot{x}_2 &= x_1,
			\end{array}
		\end{equation}
		the initial condition is $x_1=0$, $x_2=1$ and the control input is subject to the following constraints
		\begin{equation}
			-0.75 \leq u \leq 1.
		\end{equation}
		The chosen sampling time is $T_s=0.25$ seconds, the horizon length is $N=10$, and the weight matrices are $Q=Q_N=\operatorname{diag}([1.0,1.0,1.0])$ and $R=1$.
		The state trajectory obtained by each of the methods presented earlier can be seen in Figures \ref{vdp_x1} and \ref{vdp_x2}, where the similarities between them can be observed.
		
\begin{figure}
			\centering
			\includegraphics[width=1.0\linewidth]{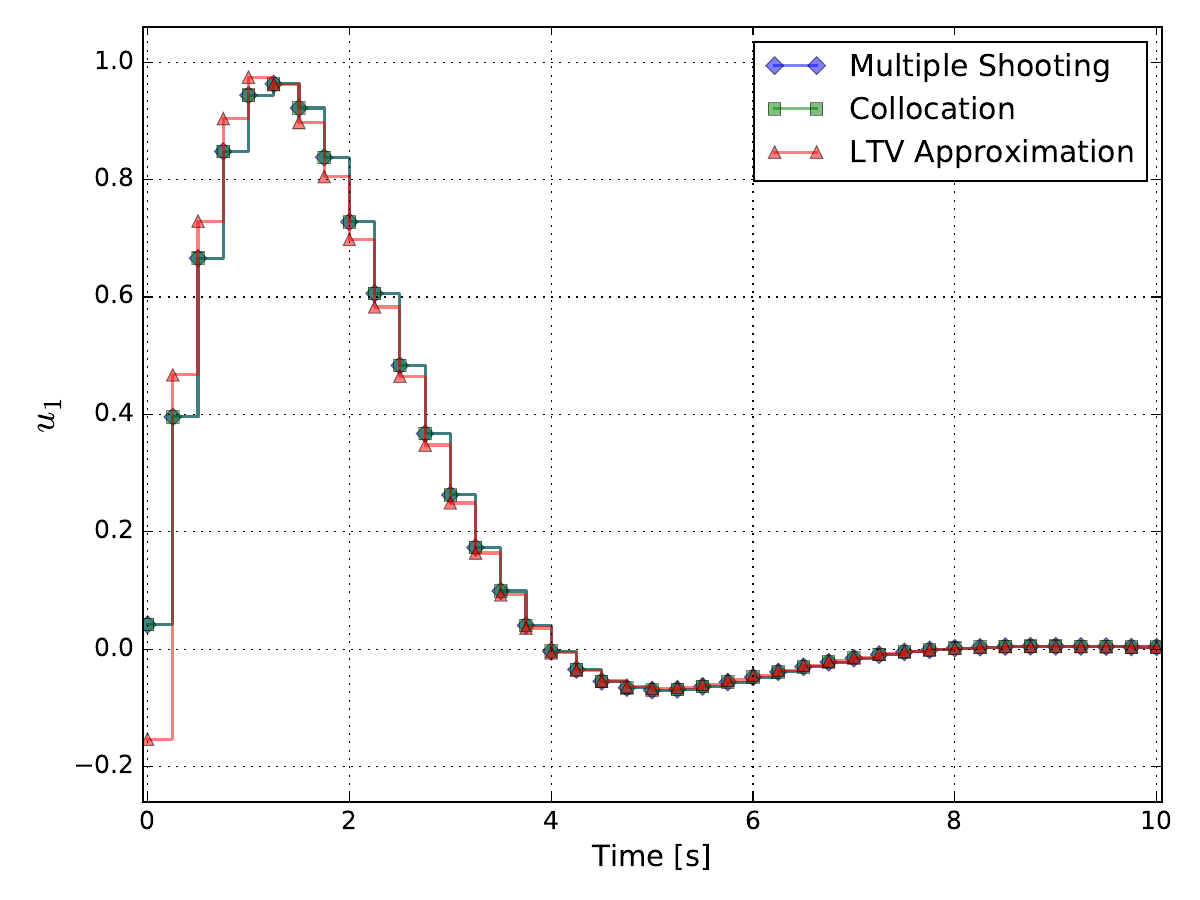}
			\caption{Control trajectory $u_1$ for the Van der Pol oscillator.}
			\label{vdp_u1}
		\end{figure}

		The obtained control input (shown in Figure \ref{vdp_u1}) is almost equal for multiple shooting and collocation, whereas the control input obtained by the successive linearization method presents some differences, mainly at the first iterations.
		
\begin{figure}
			\centering
			\includegraphics[width=1.0\linewidth]{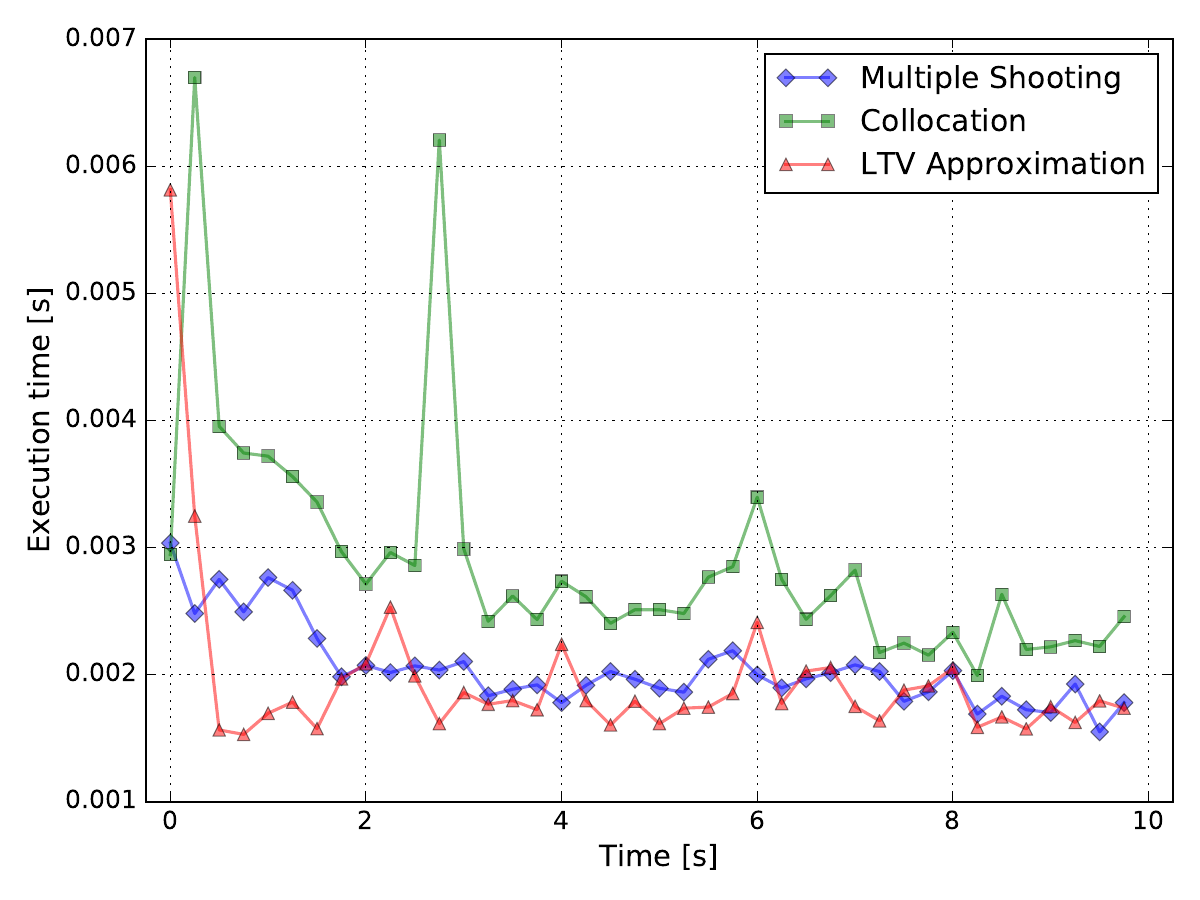}
			\caption{Execution time of each of the methods for the Van der Pol oscillator.}
			\label{vdp_proc_time}
		\end{figure}

		Finally, Figure \ref{vdp_proc_time} shows the execution time required to obtain the MPC solution at each discretization step. It can be seen that the successive linerarization method is faster than its counterparts. This is mainly because we used a QP solver and because we chose to use a receding horizon technique for the LTV model, updating only the last parameters of the corresponding LTV system and shifting the others. The mean time required to obtain a solution was $0.00205$ seconds for multiple shooting, $0.00290$ seconds for direct collocation and $0.00195$ seconds for the successive linearizations method.
		
		\subsection{Second example: Continuous stirred-tank reactor}
\begin{figure}
			\centering
			\includegraphics[width=1.0\linewidth]{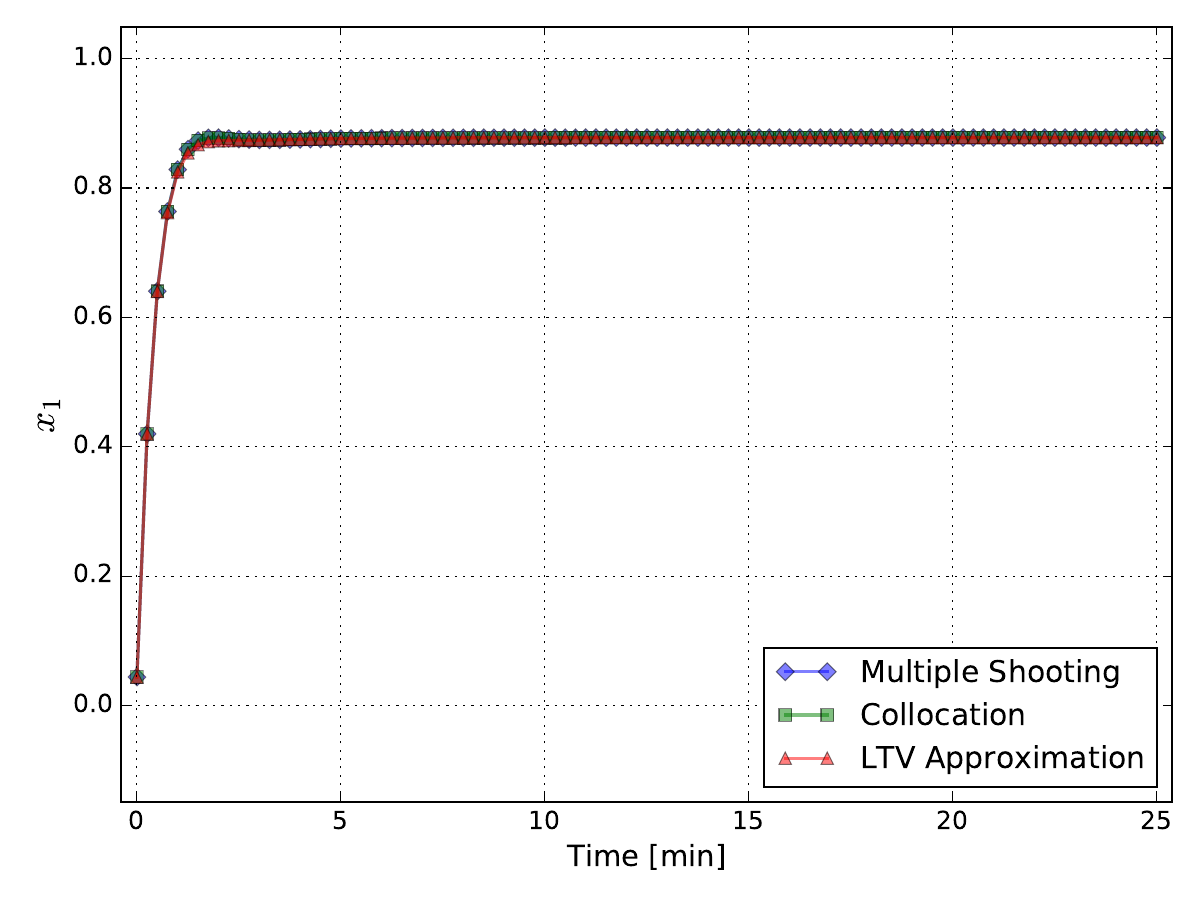}
			\caption{State trajectory $x_1$ for the CSTR.}
			\label{cstr_x1}
		\end{figure}

\begin{figure}
			\centering
			\includegraphics[width=1.0\linewidth]{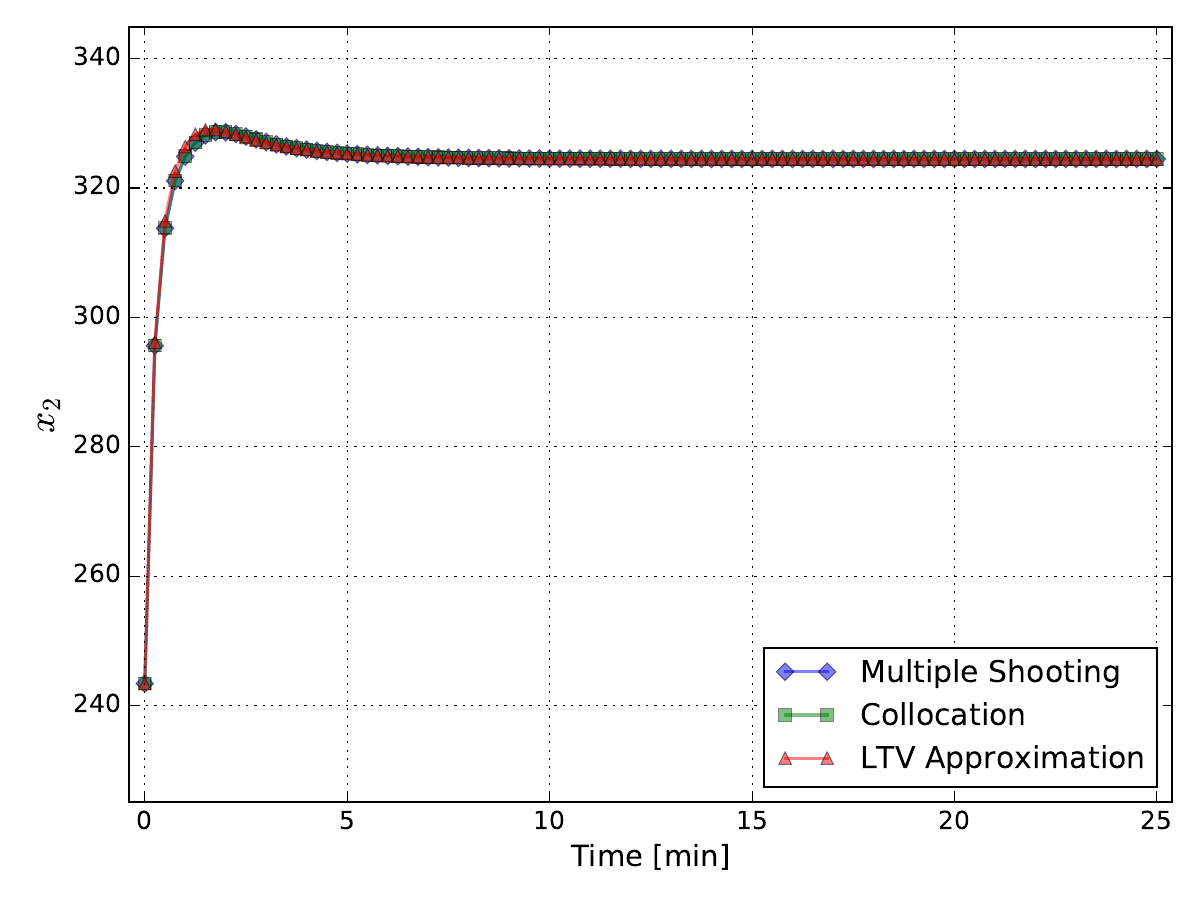}
			\caption{State trajectory $x_2$ for the CSTR.}
			\label{cstr_x2}
		\end{figure}

\begin{figure}
			\centering
			\includegraphics[width=1.0\linewidth]{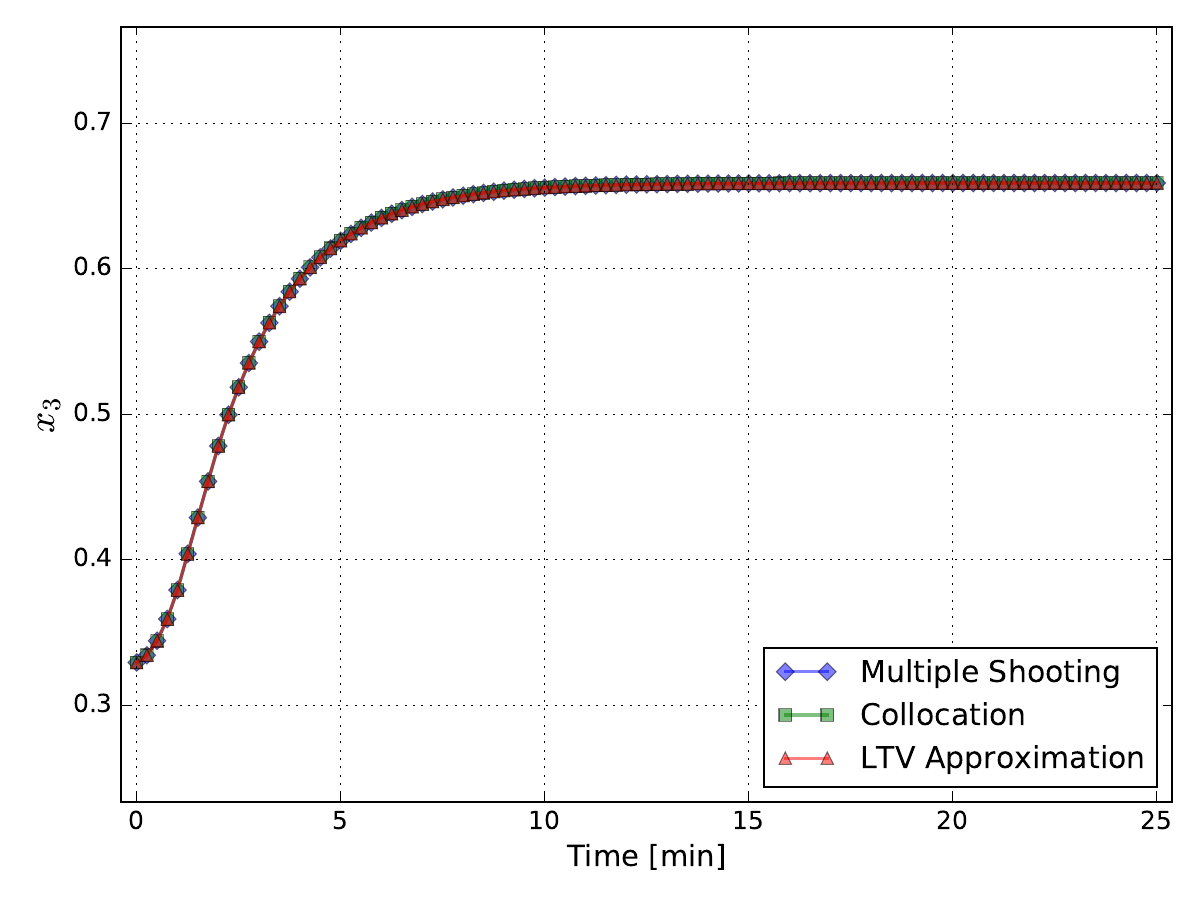}
			\caption{State trajectory $x_3$ for the CSTR.}
			\label{cstr_x3}
		\end{figure}

		In the second example, we will consider a continuous stirred-tank reactor (CSTR) taken from the work of Rawlings and Mayne \cite{Rawlings2014}. Here, an irreversible, first-order reaction, $A \to B$, occurs in the liquid phase, and the reactor temperature is regulated with external cooling. The CSTR equation is given by
		\begin{equation}
			\begin{array}{rcl}
				\dot{x}_1 &=& F_0 \frac{c_0 - x_1}{\pi r^2 x_3} - k_0 x_1 \exp(-\frac{E}{R x_2}), \\
				\dot{x}_2 &=& F_0 \frac{T_0 - x_2}{\pi r^2 x_3}
				- \frac{\Delta H}{\rho C_p} k_0 x_1 \exp(-\frac{E}{R x_2}) \\
				& & + 2 \frac{U}{r \rho Cp} (u_1 - x_2),   \\
				\dot{x}_3 &=& \frac{F0 - u_2}{\pi r^2}, \\
			\end{array}
		\end{equation}
		where $x_1$ is the molar concentration of $A$, $x_2$ is the reaction temperature in (K) and $x_3$ is the level of the tank in meters, and the controls are $u_1$ and $u_2$, the outlet flow rate in (L/min) and coolant liquid temperature, respectively. The rest of the CSTR parameters can be seen in Table \ref{tab:cstr_parameters}. In this example, we will start from the initial condition $x_1=0.04389$, $x_2=243.3730$, $x_3=0.3295$ and we wish to drive the system to the setpoint $x_1^s=0.8778$, $x_2^s=324.5$, $x_3^s=0.659$. The control input is subject to the following constraints
		\begin{equation}
			\begin{array}{rcccl}
				285 &\leq& u_1 &\leq& 315,  \\
				0.085 &\leq& u_2 &\leq& 0.115, \\
				-3 &\leq& \Delta u_1 &\leq& 3,  \\
				-0.03 &\leq& \Delta u_2 &\leq& 0.03, \\ 
			\end{array}
		\end{equation}
		and the control setpoint is expected to be $u_1^s = 300 $ and $u_2^s= 0.1$. The chosen sampling time is $T_s=15$ minutes, the horizon length is $N=15$, and the weight matrices are $Q=\operatorname{diag}([\num{6.48869d-1}, \num{4.74840d-6},   1.15132])$, $R=\operatorname{diag}([\num{2.22222d-5},200])$ and $Q_N=10 Q$.
		The state trajectory obtained by each of the methods is shown in Figures \ref{cstr_x1}, \ref{cstr_x2} and \ref{cstr_x3}, where it can be seen again the similarities between each of the reviewed methods.
		
\begin{figure}
			\centering
			\includegraphics[width=1.0\linewidth]{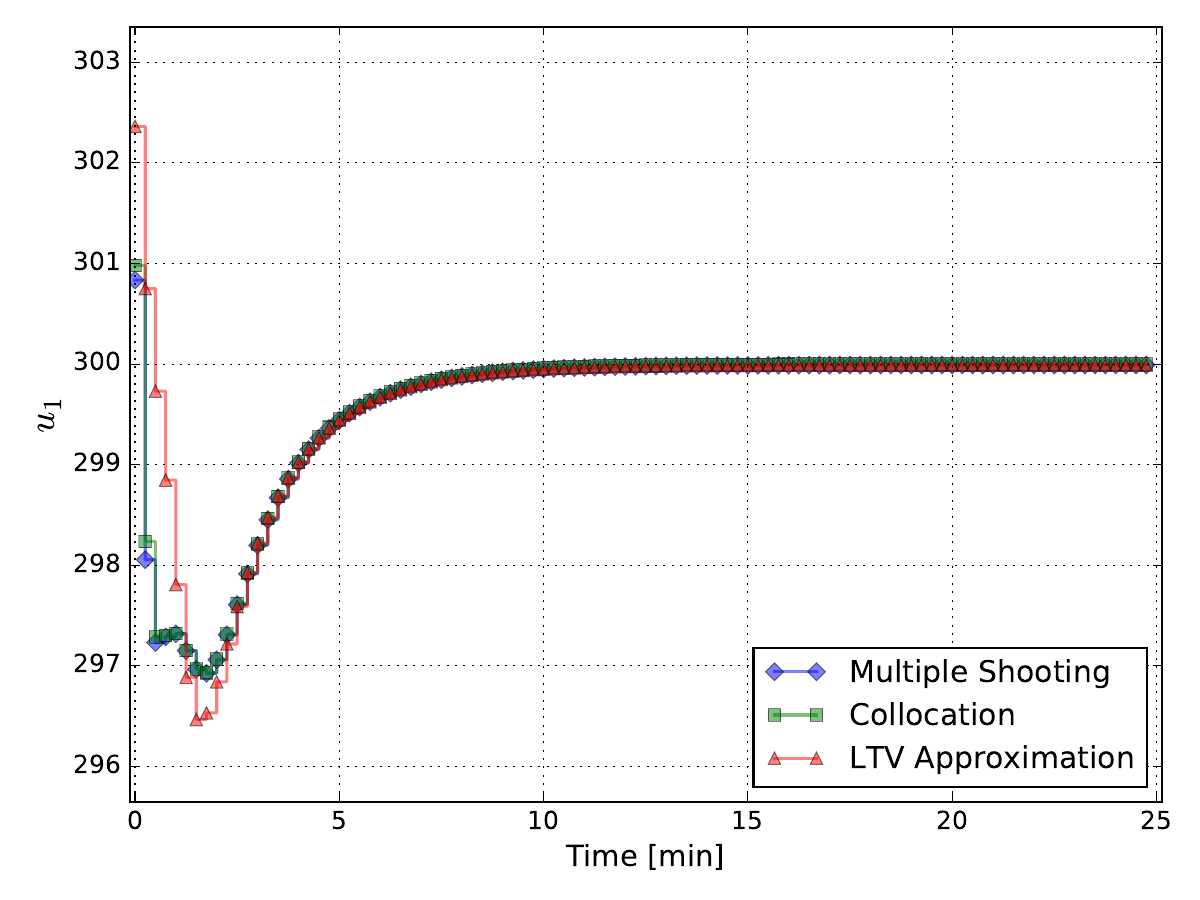}
			\caption{Control trajectory $u_1$ for the CSTR.}
			\label{cstr_u1}
		\end{figure}

\begin{figure}
			\centering
			\includegraphics[width=1.0\linewidth]{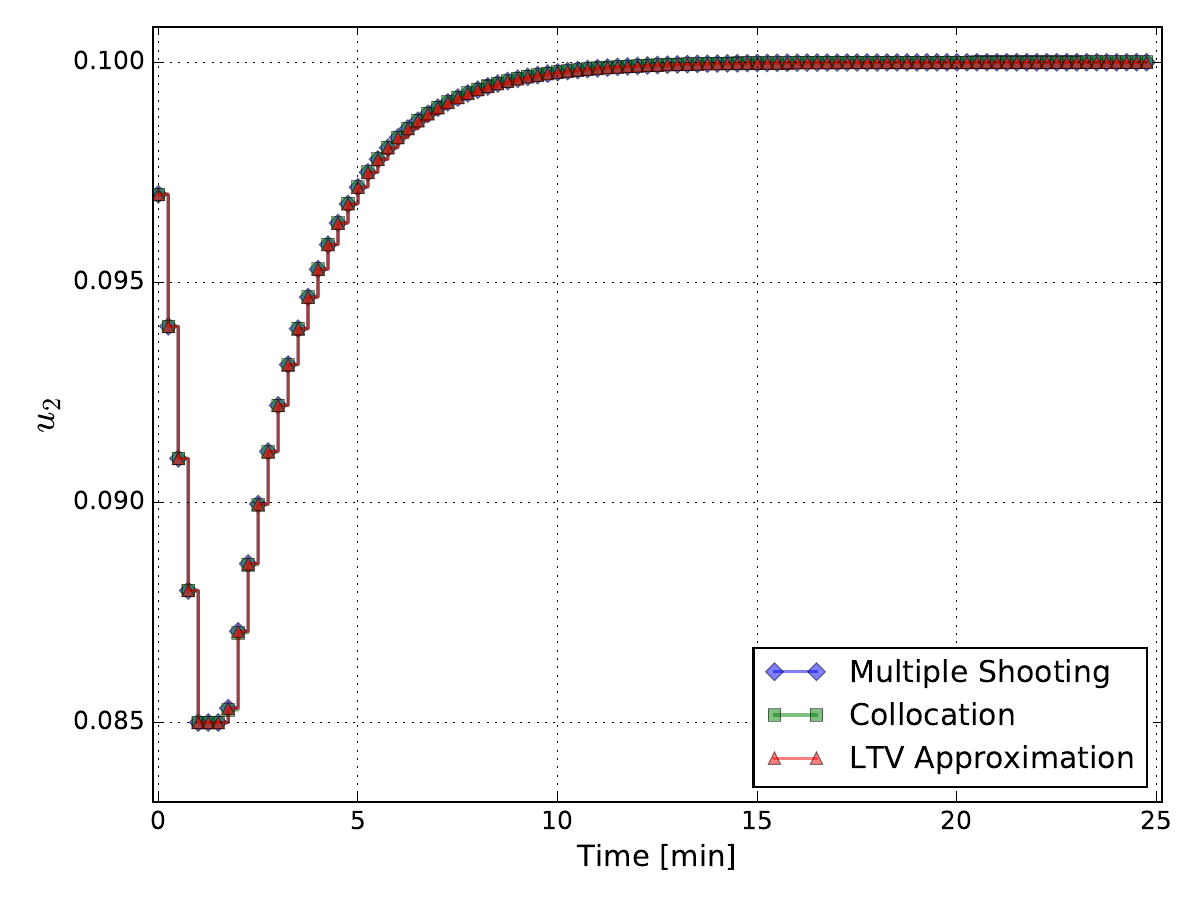}
			\caption{Control trajectory $u_2$ for the CSTR.}
			\label{cstr_u2}
		\end{figure}

		The obtained control inputs (shown in Figures \ref{cstr_u1} and \ref{cstr_u2}) are almost equal for multiple shooting and collocation, whereas --once again-- the obtained control inputs of the successive linearization method present some differences, mostly noticeable at the first iterations of $u_1$.
		
\begin{figure}
			\centering
			\includegraphics[width=1.0\linewidth]{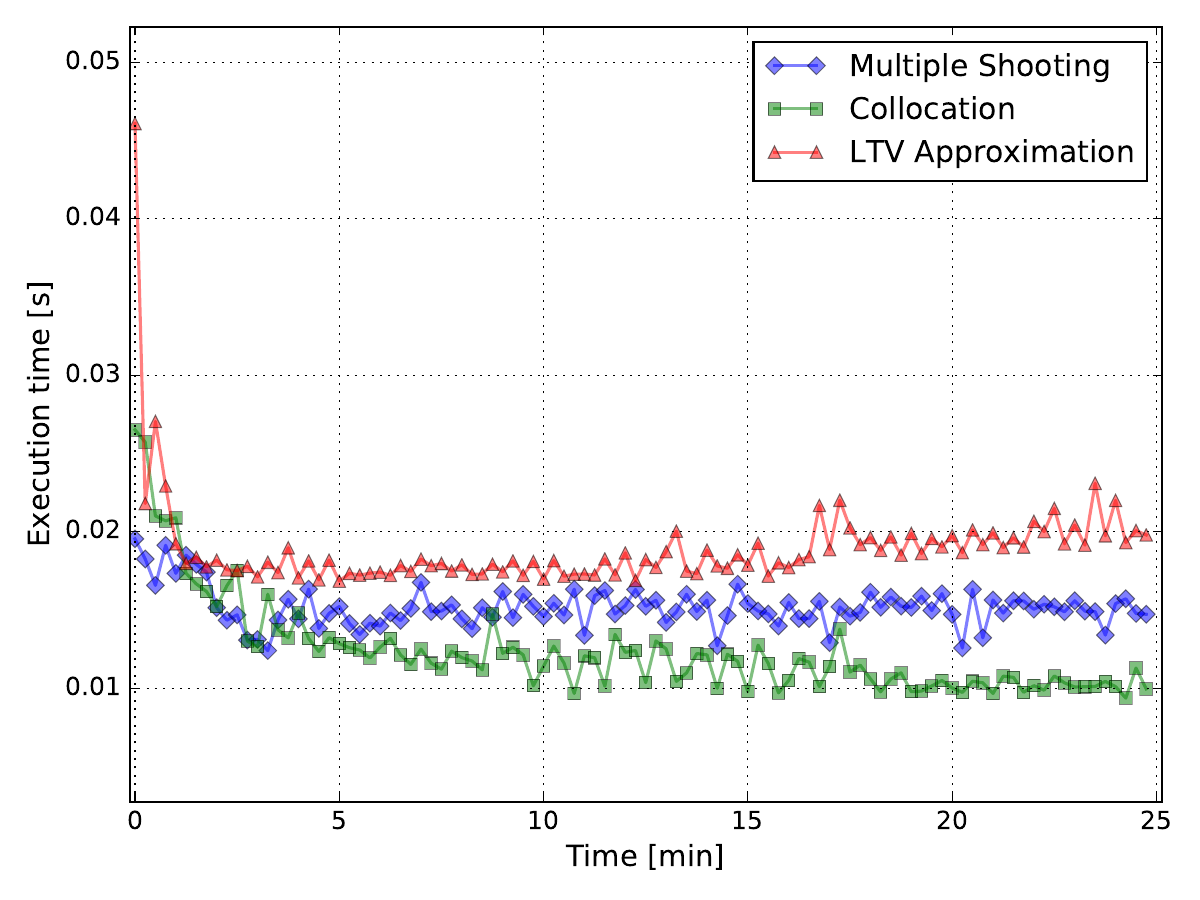}
			\caption{Execution time of each of the methods for the CSTR.}
			\label{cstr_proc_time}
		\end{figure}

		Finally, Figure \ref{cstr_proc_time} shows the execution time required to obtain the MPC solution at each discretization step. It can be seen that in this case, the collocation method is faster than its counterparts. This is mainly because the sparser NLP structure obtained by the use of collocation is better suited for the NLP solver that CasADi \cite{Andersson2013b} uses (IPOPT) and because, due to the non linearities of this system, we had to obtain a full linerization over the complete control horizon at each time step for the the LTV model of the successive linearization method. The mean time was $0.01517$ seconds for multiple shooting, $0.01236$ seconds for direct collocation and $0.01896 $ seconds for the successive linearizations method.
		
		\begin{table}[]
			\centering
			\begin{tabular}{lll} \hline
				Parameter & Nominal value & Units \\ \hline
				$F_0$ & $0.1$ & \si[per-mode=symbol]{\meter^3 \per \minute} \\
				$T_0$ & $350$ & \si[per-mode=symbol]{\kelvin} \\
				$c_0$ & $1$ & \si[per-mode=symbol]{\kilo \mol \per \meter^3} \\
				$r$ & $0.219$ & \si[per-mode=symbol]{\meter} \\
				$k_0$ & $7.2\times10^{10}$ & \si[per-mode=symbol]{\min^{-1}} \\
				$E/R$ & $8750$ & \si{\kelvin} \\
				$U$ & $54.94$ & \si[per-mode=symbol]{\kilo\joule\per\minute\per\square\meter\per\kelvin} \\
				$\rho$ & $1000$ & \si{\kilo\gram \per \meter^3} \\
				$C_p$ & $0.239$ & \si[per-mode=symbol]{\kilo\joule\per\kilo\gram\per\kelvin} \\
				$\Delta H$ & $-5\times10^4$ & \si[per-mode=symbol]{\kilo\joule \per \kilo\mol } \\ \hline
			\end{tabular}
			\caption{Parameters of the CSTR}
			\label{tab:cstr_parameters}
		\end{table}
		
		\section{Conclusion}
		\label{concl}
		
		This work shows that current state of the art NLP solvers can nowadays cope with real-time requirements, at least in small to medium scale non-linear systems. The successive linearization method performs as well and as fast as their counterparts and allows us to use analytical tools in order to analyze the stability, robustness and convergence of the controller. Besides, the possibility of using QP solvers guarantees that solution of the MPC control problem is obtained in each iteration.
		
		Each of the examples was programmed in Python and was not optimized in any way. Since CasADi has C++ interfaces, we think that there is room for improvement. The source code for the examples is available at \url{https://github.com/gmsanchez/mpc_comparison_rpic2017}.
		
		\section*{Acknowledgment}
		The authors wish to thank the \emph{Universidad Nacional de Litoral} (with CAID 501201101 00529 LI) and the \emph{Consejo Nacional de Investigaciones Cient\'ificas y T\'ecnicas} (CONICET) from Argentina, for their support. A proper recognition should also be made to the teams that created CasADi and MPCTools \cite{mpctools-casadi} and released them as open source software.
			
		\printbibliography
		
	\end{document}